# Conical Intersections, Charge Localization and Photoisomerization Pathway Selection in a Minimal Model of a Degenerate Monomethine Dye

Seth Olsen[*] and Ross H. McKenzie

*Centre for Organic Photonics and Electronics, School of Mathematics and Physics,*

*The University of Queensland, Brisbane, QLD 4072, Australia*

s.olsen1@uq.edu.au

**Abstract**

We propose a minimal model Hamiltonian for the electronic structure of a monomethine dye, in order to describe the photoisomerization of such dyes. The model describes interactions between three diabatic electronic states, each of which can be associated with a valence bond structure. Monomethine dyes are characterized by a charge-transfer resonance; the indeterminacy of the single-double bonding structure dictated by the resonance is reflected in a duality of photoisomerization pathways corresponding to the different methine bonds. The possible multiplicity of decay channels complicates mechanistic models of the effect of the environment on fluorescent quantum yields, as well as coherent control strategies. We examine the extent and topology of intersection seams between the electronic states of the dye, and how they relate to charge localization and selection between different decay pathways. We find that intersections between the $S_1$ and $S_0$ surfaces only occur for large twist angles. In contrast, $S_2/S_1$ intersections can occur near the Franck-Condon region. When the molecule has left-right symmetry, all intersections are associated with con- or dis-rotations and never with single bond twists. For asymmetric molecules (i.e. where the bridge couples more strongly to one end) then the $S_2$ and $S_1$ surfaces bias torsion about different bonds. Charge localization and torsion pathway biasing are correlated. We relate our observations to several recent experimental and theoretical results, which have been obtained for dyes with similar structure.

**I. Introduction.**

This paper is about molecules like those in Figure 1. These are *monomethine dyes*[1]. They are so called because they possess heteroaromatic rings separated by a bridge containing a single *methine*. Methine is a limit of $sp^2$ carbon where the octet rule is filled by participation in multiple Lewis structures, differing by bond alternation and formal charge distribution. A dye with more than one methine in the bridge is a *polymethine* dye, and all are *methine dyes*. Methine dyes are good dyes – an optically accessible, low lying $\pi\pi^*$ excitation is a universal feature of the chemical class, and the associated molar extinction coefficient is high[1]. Dyes like those in Fig. 1 played a crucial role in the dawn of synthetic organic chemistry[1].

This paper is about what happens in the excited states of methine dyes, and how population returns to the ground state. The fluorescence of methine dyes, and particularly monomethine dyes, is highly context-dependent. The steady-state fluorescence yield is very weak or undetectable in low-viscosity solution, but is amplified in certain environments, such as frozen glasses[2-4], biological macromolecules[5-8], or under high pressure[9]. Ultrafast fluorescence can sometimes be observed even when steady-state fluorescence cannot[3,10]; lifetimes can are lengthened in nanoconfined environments[11,12].

The increased fluorescence yield observed in several methine dyes bound to biological macromolecules[5-8] makes these molecules useful biological markers. The binding can be very specific; this specificity is useful and has led to the development of specific pairs of complementary dyes and macromolecules, called fluoromodules[8,13,14], for use in biological assays. There is a considerable dynamic range in the emission intensity of fluorogenic methine dyes – for example, the GFP chromophore motif is non-fluorescent[2], but in the native protein its fluorescence quantum yield is almost unity[15]. Apart from its large dynamic range, however, there is nothing particularly special about this behavior in the GFP chromophore. It is observed in other members of the chemical class[4-6,16].

When light is absorbed but not subsequently emitted, then there must be another channel, which successfully competes for the disposal of energy in the absorbed light. The relevant process in methine dyes is *double bond photoisomerization*. In double bond photoisomerization, the energy is dissipated into the motion of atoms and the topology of the double bond, corresponding to the relative orientation of substituents on its ends, is *erased* and *reset*.

It is natural to ask, in the context of Fig. 1, *which bond will photoisomerize?* By our definition of methine, and as highlighted in Figure 1, the location of double bonds is not the same in all the resonant structures. It is generally true that the photoisomerization of bonds within the heterocycles is not observed. This still leaves some ambiguity, though, because either bond to the methine itself will fluctuate between single and double bonding structure, in accordance with the resonance condition. For a symmetric dye, considered in isolation, these bonds are equivalent. In an asymmetric dye their equivalency will depend on the relative Lewis acid/base character of the heterocycles. The effect of the resonance splitting on the *color* of such dyes was extensively explored by Brooker[17] in the early-to-mid twentieth century.

In general, *both* of the methine bonds can undergo photoisomerization. Figure 2 displays possible photoisomerization products arising from the photoreaction of one example, the dye Thiazole Orange. Thiazole Orange becomes fluorescent upon binding to nucleic acids. There are four possible isomers. All are possible outcomes, because the reaction can proceed sufficiently to cause decay, and yet fail to populate another isomer. This is why we say that the reaction *erases and resets* the orientation of the adjoined fragments, rather than saying that it is *changed*. There are a maximum of *four* possible products available for a dye like

Thiazole Orange, because the heterocycles are different and neither is symmetric with respect to rotation about its bond to the methine.

Photoisomerization is an important photoreaction in biology[18]. It is non-destructive, leaving the molecule undamaged apart from the resetting of the orientation of the fragments. Although double bond thermal isomerization is not favorable due to high ground-state barriers, these barriers are not so high that they cannot be compensated by biomolecular interactions sufficiently to return to the original state in a timescale relevant for life. This "strategy" appears to have arisen in multiple independent contexts[18]. The best-known example is in human vision, where double bond photoisomerization is the primary event in visual signal transduction. In this case, photoisomer yields in solution[19] and protein[20] are different, clearly showing that the environment influences the photoisomer yields.

Population transfer between different Born-Oppenheimer electronic states is dominated by dynamics near configurations where potential energy surfaces touch. These *conical intersections*[21] have the local topology of a double cone when plotted over a special two-dimensional configurational subspace called the *branching plane*. The concept is illustrated in Figure 3. It is important to understand that the double-cone topology occurs *only* over the branching plane. Displacements in the subspace orthogonal to the branching plane preserve the electronic degeneracy to first order. In the higher-dimensional space, these intersections are extended connected manifolds (called *seams*) whose dimensionality is equal to the full internal dimensionality minus two. For example, the conical intersection seams mediating internal conversion in ethylene – the "hydrogen atom" of double bond photoisomerization – are *ten-dimensional manifolds!* The seams can be curved[22]. A branching plane, such as shown in Fig. 2, can be assigned to each point on the seam.

Uncertainty about which bond undergoes photoisomerization in a methine dye is compounded by the connected nature of the conical intersection seams[23,24]. It is conceivable (and, our present results suggest, *likely*) that pathways terminating in distinguishable photoisomers may actually decay by the *same seam*. Near the seam, the electronic and nuclear degrees of freedom are entangled, so that once the system is confirmed by measurement to be on the ground state electronic surface, there is statistical uncertainty introduced into its nuclear motion. In its most absurd limit, this argument destroys the very notion of distinguishable photochemical pathways. In less extreme limits, it implies that the distinguishability of different paths is tied to the distribution of *accessible* points on the intersection seam[25,26], and to details of the dynamics at different points along it[27]. In light of this, the "*which bond?*" question becomes significantly more subtle than it appears at first inspection. Understanding the position and extent of the seams relative to conceivable reaction coordinates is clearly important for a complete understanding of photoisomer yields.

The multiplicity of photoisomerization pathways available to a resonant methine dye is a problem for models of context-dependent fluorescence yield in these systems. Again, a good example is the green fluorescent protein (GFP) chromophore, where the range of fluorescence yields spans at least five orders of magnitude between solution and protein environments. This has been highlighted in molecular mechanics studies[28-31], which suggested that steric hindrance by the protein was *insufficient* to stop decay by all possible pathways. Nevertheless, the high observed yield of the protein implies that they *are* stopped by a mechanism that is, as yet, poorly understood.

The non-destructive nature of double bond photoisomerization, and its biological importance, has made it a useful as a testing ground for coherent chemical control techniques. Some model systems which have been used in this field are symmetric monomethine dyes – specifically, 1,1'-DiEthyl-3,3'-Thiacyanine[32] (NK88, Fig. 1), 1,1'-DiEthyl-4,4'-Cyanine(1144C)[33] and 1,1'-DiEthyl-2,2'-Cyanine[34] (1122C, Fig 1.). NK88 and 1144C are, to within a methyl-ethyl substitution, the symmetric parents[17] of the fluorogenic dye Thiazole Orange (Fig. 1)[16]. These dyes have been the subject of several spectroscopic investigations[33,35-40]. In the cases of NK88[40] and 1122C[35], evidence was cited for decay on different timescales – a fast timescale, (1-200fs for 1122C[35], and 1-2ps for NK88[40]), and a slower one (~4ps for 1122C, and 9-11ps for NK88). Different interpretations were given for the distinct timescales in either case. In the experiments on 1122C, it was explained by invocation of a proposed[12] seam of intersection near the Franck-Condon point (ground state minimum), whose access did not require substantial twist, and the slow timescale by more conventional torsional motion. In the NK88 experiments, the slow timescale was interpreted as indicative of a distinct symmetry-preserving pathway[41], while the fast timescale was interpreted as decay by a direct (non-symmetry preserving) torsional pathway.

We wish to understand how photoisomerization yields are controlled, both in the case of environmental fluorescence tuning and in coherent control, but this problem is complicated by the fact that there are *multiple* pathways for photoisomerization, as mandated by the resonance in Fig 1. If the fluorescence yield increases due to a constraining environment, it is reasonable to say that every possible decay channel represents an additional constraint that the environment must fulfill in order to prevent fluorescence quenching. In the case of coherent control, it would be useful to determine if distinct channels can be *independently addressed*. In either case, we must understand *if* and *how* pathways can be distinguished, and this requires understanding the seams themselves. Figure 4 schematically outlines how the position of the seams can affect the relevant pathways for decay in a multi-state system, and how the multi-state and multimode character could give rise to different dynamics following excitation to different states.

In this paper we will use a simple physical model to better understand the intersection seams of a general monomethine dye. This effort is designed to complement our ongoing effort to examine *specific cases* by the apparatus of computational quantum chemistry[42-45]. We hypothesize, on the basis of the widespread character of the phenomena addressed, that there is an underlying physics at work that is *transferrable, general and that can be understood*. Of course, this generality makes the problem difficult, so we have to start very simply.

In Section II, we briefly describe a family of self-consistent solutions to the electronic structure of methine dyes. These can be cast into a complete active space valence bond form, and form a heuristic basis for our model. In Section III we propose a model Hamiltonian for the interaction of group-localized valence-bond states during the photoisomerization of a hypothetical monomethine dye. In section IV, we loosely identify the parameters of the model to geometric and chemical characteristics of such a dye, using a simple "toy" molecular frame. In section V we describe analytic expressions for the eigenenergies of the model, and in section VI we derive equations that we will use to locate the intersection seams. Sections VII-XIII describe a series of specific and relevant observations relating to intersections between different states in the model. Section XIV includes a discussion of how our results relate to current knowledge of, and recent experiments on, double bond photoisomerization in monomethine dyes. We conclude in Section XV.

**II. Electronic Structure of Monomethine Dyes**

The model Hamiltonian that we describe here is conceptually based upon a common family of solutions to the state-averaged complete active space self-consistent field[46] problem for monomethine dyes. This family of solutions has a common structure when expressed in a localized orbital representation, summarized in Figure 5. Solutions with this structure are easily obtained for the SA-CASSCF problem with either four or three electrons distributed over three orbitals for the two classes of monomethine dye with two heterocyclic rings: diarylmethine dyes and monomethine cyanine dyes. The two classes are exemplified in Fig 5 by Michler's Hydrol Blue (left) and 1,1'-Dimethyl-2,2'-

Cyanine (right). A key difference is that the diarylmethine heterocycles are aromatic in their reduced state, while the cyanine heterocycles are aromatic in their oxidized state.

For either class of methine dyes, there are three orbitals and six singlet configurations defined over the orbitals. Three of the configurations are "covalent" and three are "ionic", as in Fig. 5. If the energies of the covalent and ionic states are separated, then it is reasonable to define an effective model space indexed by the covalent states alone. We have previously exploited this to formulate a diabatic picture of the photoisomerization in the green fluorescent protein chromophore[42]. In the reduced space, each degree of freedom is equivalent to a covalent-ionic contracted pairing state. Our model Hamiltonian is a parametric approximation to the Hamiltonian defined on the contracted model space.

**III. Model Hamiltonian.**

In this section, we propose a model Hamiltonian for an idealized degenerate monomethine dye. The model is justified by the valence-bond structure of the self-consistent solutions described above. The specific parameterization is based on the established principle that the interaction of valence-bond states can be approximated as orbital overlaps[47]. This idea is at the heart of chemical thought[48,49]. The strategy of parameterizing model Hamiltonian matrices based on computational and/or experimental results to describe reactivity has been employed successfully by Bernardi, Robb and coworkers[50], by Warshel and Weiss[51], by Chang and Miller[52], and by Malrieu and coworkers[53]. Excited-state and photochemical models following a similar strategy have been applied by Bernardi, Robb and coworkers[54], Bearpark and coworkers[55], by Said, Maynau and Malrieu[56], by Ben-nun and coworkers[57], by Burghardt and Hynes[58], and by Domcke and coworkers[59].

We begin by writing down our model Hamiltonian as the 3 x 3 matrix (1).

$$\mathbf{H} \equiv \begin{pmatrix} H_{ll} & H_{lb} & H_{lr} \\ \vdots & H_{bb} & H_{br} \\ \cdot\cdot & \ldots & H_{rr} \end{pmatrix} = \begin{pmatrix} E_0 & -\zeta \cos(\theta_L) & -\xi \cos(\theta_R - \theta_L) - \psi \cos(\theta_R + \theta_L) \\ \vdots & E_0 & -\zeta^{-1} \cos(\theta_R) \\ \cdot\cdot & \ldots & E_0 \end{pmatrix} \quad (1)$$

Where the variables $\theta_L$ and $\theta_R$ are angles (in radians) and the parameters $\xi, \psi$ and $\zeta$ are real, positive numbers. We can, by invocation of a trigonometric identity, write the model in an equivalent alternate form (2).

$$\mathbf{H} = \begin{pmatrix} E_0 & -\zeta \cos(\theta_L) & -(\xi+\psi)\cos(\theta_L)\cos(\theta_R) - (\xi-\psi)\sin(\theta_L)\sin(\theta_R) \\ \vdots & E_0 & -\zeta^{-1} \cos(\theta_R) \\ \cdot\cdot & \ldots & E_0 \end{pmatrix} \quad (2)$$

Most of the remainder of this paper is concerned with the behavior of the eigenstates of **H** as the angles are varied for set fixed values of the three parameters $\xi, \psi,$ and $\zeta$.

**IV. Interpretation of Angles and Parameters in Terms of a Molecular "Toy" Model.**

In this section, we will offer an interpretation of the angles ($\theta_L$ and $\theta_R$) and parameters ($\xi, \psi,$ and $\zeta$) in terms of a hypothetical molecular frame. Consider the "toy" model at the bottom of Figure 6. The "toy" consists of three p orbitals arranged on a triangle. We will show that the off-diagonal matrix elements of **H** can be expressed simply in terms of the overlaps between the p orbitals, subject to some constraints on their relative orientation.

Suppose that the bridge orbital *b* at the apex of the triangle is fixed so that its long axis is perpendicular to the plane of the triangle. Also, the orbitals on either end (*l* and *r*) rotate rigidly about their bonds to *b*, so that their long axes are always perpendicular to the bond. We can express the overlaps between the p orbitals by decomposing their total overlap into π and σ components of the overlap multiplied by direction cosines[60]. Doing this we obtain the overlap formulas (3-5).

$$S_{br} = -S_\pi(r_R)\cos(\theta_R) \tag{3}$$

$$S_{lb} = -S_\pi(r_L)\cos(\theta_L) \tag{4}$$

$$S_{lr} = (S_\pi(r_B))\cos(\theta_L)\cos(\theta_R) \\ + (S_\sigma(r_B)\sin(\phi_L)\sin(\phi_R) + S_\pi(r_B)\cos(\phi_L)\cos(\phi_R))\sin(\theta_L)\sin(\theta_R) \tag{5}$$

We can use the same trigonometric identity used to equate (1) and (2) in reverse to rewrite (5) as (6).

$$S_{lr} = (S_\pi(r_B) + S_\sigma(r_B)\sin(\phi_L)\sin(\phi_R) + S_\pi(r_B)\cos(\phi_L)\cos(\phi_R))\cos(\theta_R - \theta_L) \\ + (S_\pi(r_B) - S_\sigma(r_B)\sin(\phi_L)\sin(\phi_R) - S_\pi(r_B)\cos(\phi_L)\cos(\phi_R))\cos(\theta_R + \theta_L) \tag{6}$$

The matrix elements of **H** have the same dependence on the angles $\theta_L, \theta_R$ as the overlaps between the orbitals in the "toy". Now, we can equate coefficients of the trigonometric functions of $\theta_L$ and $\theta_R$ and write expressions (7-9).

$$\xi = \frac{S_\pi(r_B) + S_\sigma(r_B)\sin(\phi_L)\sin(\phi_R) + S_\pi(r_B)\cos(\phi_L)\cos(\phi_R)}{\sqrt{S_\pi(r_L)S_\pi(r_R)}} \tag{7}$$

$$\psi = \frac{S_\pi(r_B) - S_\sigma(r_B)\sin(\phi_L)\sin(\phi_R) - S_\pi(r_B)\cos(\phi_L)\cos(\phi_R)}{\sqrt{S_\pi(r_L)S_\pi(r_R)}} \tag{8}$$

$$\zeta \propto \sqrt{\frac{S_\pi(r_L)}{S_\pi(r_R)}} \tag{9}$$

Alternatively, as regards the alternate form of **H** given in (2), we could take sums and differences of (7) and (8) to obtain (10) and (11).

$$\xi + \psi = \frac{S_\pi(r_B)}{\sqrt{S_\pi(r_L)S_\pi(r_R)}} \quad (10)$$

$$\xi - \psi = \frac{S_\sigma(r_B)\sin(\phi_L)\sin(\phi_R) + S_\pi(r_B)\cos(\phi_L)\cos(\phi_R)}{\sqrt{S_\pi(r_L)S_\pi(r_R)}} \quad (11)$$

We now have all that we need to establish the relationship between **H** and the overlaps. This relationship is expressed in (12).

$$H_{i \ne j} = \frac{-S_{ij}}{\sqrt{S_\pi(r_L)S_\pi(r_R)}} \quad (12)$$

So that the interaction matrix elements between the different orbitals are proportional to the overlap, but normalized by the geometric mean of the bridge-end overlaps at zero twist. We can identify $\xi$ with the relative magnitude of the left-right coupling when the molecule is planar, $\psi$ with the magnitude of the same coupling when both bonds are rotated through $90^0$, and $\zeta$ with the ratio of coupling of the left and right ends, respectively, to the bridge. Alternatively, we could interpret $\xi+\psi$ as the coefficient of coupling modulated by the antisymmetric angle coordinate $\theta_L$-$\theta_R$ and $\xi$-$\psi$ as the coefficient of coupling modulated by the symmetric angle coordinate $\theta_L+\theta_R$.

To make the interpretation even more concrete, let us assume that the overlaps can be written as a product of an exponential and a polynomial in a generalized distance, as in (13).

$$S_\pi(r_{ab}) = S_{2p\pi}(p_{ab}) = e^{-p_{ab}} P_{2p\pi}(p_{ab}) \tag{13}$$

$$S_\sigma(r_{ab}) = S_{2p\sigma}(p_{ab}) = e^{-p_{ab}} P_{2p\sigma}(p_{ab}) \tag{14}$$

Here $P_{2p\pi}$ and $P_{2p\sigma}$ are polynomials tabulated by Mulliken et al.[60], and $p_{ab}$ is the 'effective distance' between atoms $a$ and $b$, given by (15).

$$p_{ab} \equiv \frac{(\mu_a + \mu_b) r_{ab}}{2 a_H} \tag{15}$$

Where $\mu_a$ and $\mu_b$ are parameters reflecting the "atomic type", $r_{ab}$ is the real distance between the atoms, and $a_H$ is the Bohr radius.

If we substitute (13) and (14) into (7), (8) and (9) and factor out the exponentials, we arrive at proportionality expressions that relates the parameters $\xi$, $\psi$ and $\zeta$ to the distances $r_L$, $r_R$ and $r_B$ (16-18).

$$\xi \propto e^{-(r_B - \frac{r_L + r_R}{2})} \tag{16}$$

$$\psi \propto e^{-(r_B - \frac{r_L + r_R}{2})} \tag{17}$$

$$\zeta \propto e^{-(\frac{r_L - r_R}{2})} \tag{18}$$

These expressions state that the parameters $\xi$, $\psi$ and $\zeta$ reflect overlap modulation by bond length and angle in a hypothetical methine unit. The parameters $\xi$ and $\psi$ express the modulation of the overlap by extension of $r_B$ at the expense of $r_L$ and $r_R$, which is a symmetric stretch-bend distortion, and $\zeta$ expresses modulation of the coupling by an antisymmetric stretch.

In practice, we are going to consider *only* the angles $\theta_L$ and $\theta_R$ as "dynamical variables" in this paper, and will not be making explicit references to the distances, real or effective, between orbitals. This is in line with the approximate nature of the connection between the parameters $\xi$, $\psi$ and $\zeta$ and the geometry of any real dye. Clearly, for the dyes shown in Fig. 1, the "effective distance" would, at least, have to take into account the chemical identity of the heterocyclic nuclei, and the resulting expressions would quickly get cumbersome. We expect that the angular dependence on $\theta_L$ and $\theta_R$ would remain in similar form, however, as long as the orbitals are of $\pi$ type, as in Fig. 5. In situations where the twisting is much slower than the skeletal vibrations, we can consider $\xi$, $\psi$ and $\zeta$ as representative of a distribution of geometries for the methine unit.

Most of the figures here will contain plots of intersections or regions containing intersections over a plane spanned by the angles $\theta_L$ and $\theta_R$. This "torsion plane" is a form of periodic lattice, with periodicity defined by the periodicity of the angles $\theta_L$ and $\theta_R$, and the invariance of the cosine function under negation of its argument. We will speak occasionally of a "unit cell" within the plane, by which we mean a particular periodic image – a square patch of the lattice with length $\pi$ on a side. The torsion plane is summarized in Figure 7, along with the chemical meaning of certain directions within the plane. Note that our definition of "dis-" and "con-rotatory" directions is dependent on a "handedness convention" wherein we use different hands to determine the positive direction of rotation for angles about different bonds. The reason for this convention is simple – when discussing the model amongst ourselves, it has proven useful to use both hands.

In our model, we have assumed that the diabatic states are degenerate. This is not the case for any of the dyes in Fig. 1, according to our ongoing *ab initio* studies. Even for the symmetric monomethine dyes, there will be a splitting between the ring states and the bridge, the sign of which depends on whether one is dealing with a dye of arylmethine or cyanine type (see Fig. 5). We have chosen to examine the degenerate case as a reference case, to which we will refer when we address more complicated models at a later stage. In general, the location of conical intersections between two states requires zero coupling *and* zero splitting[61], and our constraint guarantees the latter. We expect that the degeneracy will accentuate the presence of intersections in the model. Our continuing investigations suggest that the inclusion of reasonable splitting does not significantly change our most important results.

Our supposition that the bridge p orbital in our "toy" remain perpendicular to the plane does not allow for pyramidal distortion of the bridge. Likewise, the supposition that the p orbitals on the ends are orthogonal to the bonds does not allow for pyramidalization of the methine-adjoining atoms on the heterocycles. Pyramidalization and torsion motions are not independent. The explicit inclusion of pyramidalization may not allow the factorization of the overlap in the toy into distinct σ and π components multiplied by direction cosines, as we have done in eqns. (3)-(5). Pyramidalization is known to occur at $S_0/S_1$ conical intersections in models of fluorescent protein chromophores[43-45,62]. Still, it is possible that the effects of pyramidalization could be *approximately* embedded in a suitable choice of the parameters of the model.

**V. Analytic Formulas for Eigenenergies and Eigenvectors**

In this section we review analytic formulas for the eigenvalues and eigenvectors of a 3 x 3 matrix, which we will use to probe the model. The formulas for the eigenvalues of a 3 x 3 matrix are described by Cocoliccio and Viggiano[63], and we roughly follow their exposition for the real, symmetric case.

To begin, we define three quantities, which are generalized moments (first and higher-order traces, and determinants) of the matrix **H**.

$$b = -Tr\mathbf{H} \tag{19a}$$

$$c = \frac{1}{2}\left[(Tr\mathbf{H})^2 - Tr(\mathbf{H}^2)\right] \tag{19b}$$

$$d = -Det\mathbf{H} \tag{19c}$$

With these, we define the accessory quantities $v$ and $u$ and an angle $\Theta$ (20-22).

$$v = \frac{1}{54}\left(2b^3 - 9bc + 27d\right) \tag{20}$$

$$u = -\frac{1}{9}\left(b^2 - 3c\right) \tag{21}$$

$$\cos\Theta = -\left(\frac{v}{u\sqrt{|u|}}\right) \quad (23)$$

The requirement that the matrix be real and symmetric dictates that $u<0$, and that $-v^2 \leq u^3$.

The eigenenergies $\varepsilon$ of **H** can be expressed completely, up to a shift, in terms of $u$ and $v$.

$$z_{i \in \{0,1,2\}} = \varepsilon_i - \frac{b}{3} = \varepsilon_i - E_0 \quad (24)$$

$$z_0 = -2\sqrt{u}\cos\left(\frac{\Theta}{3}\right) \quad (25)$$

$$z_1 = \sqrt{|u|}\left[\cos\left(\frac{\Theta}{3}\right) - \sqrt{3}\sin\left(\frac{\Theta}{3}\right)\right] \quad (26)$$

$$z_2 = \sqrt{|u|}\left[\cos\left(\frac{\Theta}{3}\right) + \sqrt{3}\sin\left(\frac{\Theta}{3}\right)\right] \quad (27)$$

In keeping with the usual terminology of organic photochemistry, we will refer to the states with eigenenergies $\varepsilon_0$, $\varepsilon_-$ and $\varepsilon_+$ as $S_0$, $S_1$, and $S_2$. Formulas for the (unnormalized) eigenvectors are given in (28)[64].

$$\vec{v}_{i \in \{0,1,2\}} = \begin{pmatrix} -(\varepsilon_i - H_{22})H_{13} - H_{12}H_{23} \\ -(\varepsilon_i - H_{11})H_{23} - H_{21}H_{13} \\ -(\varepsilon_i - H_{11})(\varepsilon_i - H_{22}) + H_{12}^2 \end{pmatrix} \quad (28)$$

The populations of the diabatic states in the eigenstates can be obtained in the usual way, but taking the squares of amplitudes. The identification of the diabatic states with the covalent valence bond functions in Fig. 5 allows us to take these populations as a probability distribution for the location of the excess charge.

## VI. Full and Partial Location of the Intersection Seams

The conditions for $S_1/S_0$ intersection and $S_2/S_1$ intersection can be obtained by setting (25) equal to (26), or (26) equal to (27), respectively. If we do this, we find that $S_0/S_1$ intersections occur when $\Theta=\pi$ and that $S_2/S_1$ intersections will occur when $\Theta=0$. Substituting this into (23) we easily obtain that the conditions for degeneracy can be expressed as (29).

$$v \mp u\sqrt{|u|} = 0 \qquad (29)$$

Where the minus sign indicates a $S_1/S_0$ intersection and the plus sign indicates a $S_2/S_1$ intersection (again, $u$ is guaranteed negative). A 3-State intersection corresponds to a trivial solution where both $v$ and $u$ are zero. All three circumstances can be expressed in a single equation if we square (29) to obtain (30).

$$v^2 + u^3 = 0 \qquad (30)$$

Equation (30) is a sixth order polynomial in the matrix elements of **H**. Although its solutions would, in principle, provide a complete characterization of degeneracies in the model, these solutions are not straightforward to obtain. We will not attempt to find a general solution to either (29) or (30) in this paper.

In lieu of attempting to solve (29) or (30) directly, we will be using two other methods to characterize the extent and location of intersections in the model. To illustrate the first, we point out that eqn. (29) also implies that (31) will hold in "an open neighborhood" of the degeneracy seams, for a suitable small number $\delta$.

$$\left| u\sqrt{|u|} - |v| \right| < \delta \tag{31}$$

In many of the figures in this paper, we will be using eqn. (31) to *partially* locate the seams within open neighborhoods, rather than attempting to solve (29) or (30) directly.

The real, symmetric character of **H** implies $u \leq 0$ and $-v^2 \leq u^3$. Given this, the implication of (29) is that *v must be negative when $S_1$ and $S_0$ intersect, positive when $S_2$ and $S_1$ intersect, and u and v must simultaneously vanish when all three eigenenergies are degenerate.* These conditions, which are *necessary – but not sufficient – conditions* for a given pair (or triple) of states to intersect in any region of the combined angle-parameter space, are summarized in eqns. (32).

$S_1/S_0$ Intersection:  $v < 0$ \hfill (32a)

$S_2/S_1$ Intersection:  $v > 0$ \hfill (32b)

3-State Intersection: $v = 0$ \hfill (32c)

Eqns (32) are very useful, because *v* is given by a 2nd order polynomial in the matrix elements, whereas equation (30) is a 6th order polynomial in these elements! Figure 8 summarizes the relationship between *u, v* and the intersection seams for a general case. Figure 9 demonstrates the use of eqn. 31 to visualize the intersection seams in the simple limit where $\psi=0$ and $\zeta=1$.

Having established the importance of *u* and *v* as tools to query the location and topology of intersection seams within the model, it would be useful to write them explicitly in terms of the parameters and angles. Substituting (1) or (2) into (20), we can obtain explicit expressions (33) and (34) for *v* in terms of the parameters and angles.

$$v = (\xi \cos(\theta_L - \theta_R) + \psi \cos(\theta_L + \theta_R))\cos(\theta_L)\cos(\theta_R) \tag{33}$$

$$v = ((\xi + \psi)\cos(\theta_L)\cos(\theta_R) + (\xi - \psi)\sin(\theta_L)\sin(\theta_R))\cos(\theta_L)\cos(\theta_R) \tag{34}$$

We can do the same for *u*, obtaining (35) and (36).

$$u = \frac{-1}{3}\left(\zeta^2 \cos^2(\theta_L) + \zeta^{-2} \cos^2(\theta_R) + (\xi \cos(\theta_L - \theta_R) + \psi \cos(\theta_L + \theta_R))^2\right) \tag{35}$$

$$u = \frac{-1}{3}\left(\zeta^2 \cos^2(\theta_L) + \zeta^{-2} \cos^2(\theta_R) + ((\xi + \psi)\cos(\theta_L)\cos(\theta_R) + (\xi - \psi)\sin(\theta_L)\sin(\theta_R))^2\right) \tag{36}$$

We will now summarize the main points of this section: **1.** The sign of *v* (eqns. 33 & 34) provides a *necessary* condition on the location of intersections between a given pair (or triple) of states. **2.** *Sufficient* conditions require information about *u* as well (eqns. 35 & 36). Complete location of the seam requires solution of equations (29) and (30), which is difficult. **3.** *Partial* localization of the seam is possible by finding regions where equation (31) holds for a sufficiently small number $\delta$. The rest of the paper describes the application of these ideas to the model, and highlights chemical ramifications that emerge.

**VII. $S_1/S_0$ intersections can only occur at twisted geometries.**

In the previous section, we established that, for any point in the model space specified by values of the parameters ($\xi$, $\psi$ and $\zeta$) and angles ($\theta_L$ and $\theta_R$), the states that *can* intersect at is determined by the sign of *v* (eqns. 33 & 34). There are two points to make regarding (33) and (34). First, *v* does not depend on $\zeta$, because wherever $H_{lb}$ appears, it is multiplied by $H_{br}$. These matrix elements have $\zeta$ in the numerator and denominator, respectively, so the $\zeta$ dependence cancels. This means that the asymmetry of the coupling between the bridge and the left and right molecular fragments has a limited effect on the character of the intersections. Secondly, if both $\xi$ and $\psi$ are positive (as assumed), then $(\xi+\psi)cos^2\theta_L cos^2\theta_R$ is also positive, so we can divide $v$ by this and obtain a quantity with the same sign as $v$. Therefore, the necessary conditions for $S_1/S_0$, $S_2/S_1$ or 3-State intersections can be expressed in a simplified form that only depends on the angles $\theta_L$ and $\theta_R$ and a *single affine parameter* $\frac{\xi-\psi}{\xi+\psi}$.

$$S_1/S_0 \text{ Intersection: } 1+\frac{\xi-\psi}{\xi+\psi}tan(\theta_L)tan(\theta_R)<0 \quad (37a)$$

$$S_2/S_1 \text{ Intersection: } 1+\frac{\xi-\psi}{\xi+\psi}tan(\theta_L)tan(\theta_R)>0 \quad (37b)$$

$$3\text{-State Intersection: } 1+\frac{\xi-\psi}{\xi+\psi}tan(\theta_L)tan(\theta_R)=0 \quad (37c)$$

The point that we wish to make here is that (37) implies $S_1/S_0$ intersections only occur in a restricted range of geometries that are characterized by a high degree of bridge twist. The accessible area of the torsion plane in which such intersections can occur is greatest when the affine parameter $\frac{\xi-\psi}{\xi+\psi}=\pm 1$. It is least for $\frac{\xi-\psi}{\xi+\psi}=0$, in which case $S_1/S_0$ intersections only occur as part of 3-state intersections. The situation is summarized in Figure 10.

It is interesting that we only observe $S_1/S_0$ intersections at highly twisted geometries, because this conflicts *prima facie* with a recent proposal that there is an accessible seam of conical intersection close to the Franck-Condon region of methine dyes. The proposal was based on nonadiabatic surface-hopping simulations of a trimethine streptocyanine dye[12]. This proposal has recently been invoked to explain the observation of an extremely fast pathway in 1,1'-Diethyl-2,2'-Cyanine[35], a symmetric monomethine cyanine dye. Further work is needed to understand the origin of this apparent conflict, and find possibilities for its resolution.

## VIII. When the coupling to the bridge is the same for both ends, all intersections lie along con- or dis-rotatory coordinates

We find that when $\zeta=1$, all intersections in the model occur along dis- or con-rotatory twist coordinates. Setting $\zeta=1$ means that the bridge-end couplings are symmetric, and should be representative of a symmetric monomethine dye. In the context of our "toy" molecule discussed in Section II, variation in $\zeta$ can be thought of as distortion along an antisymmetric stretching coordinate. Of course, similar modulation of $\zeta$ could also be achieved by changing the identity of the end groups, because this would change the 'atomic type', as embodied in the $\mu$'s in equation (15).

The effect of varying $\zeta$ on the structure of the intersection seams is graphically summarized in Figure 11, for the limit where $\psi=0$.

When $\zeta=1$, the intersection seam (both $S_1/S_0$ and $S_2/S_1$ branches) is restricted to lie along con- or disrotatory lines in the torsion plane, which intersect at the point $(\pi/2,\pi/2)$, where a 3-State intersection occurs. As shown, deviation of $\zeta$ from 1 will lead to a distortion of the seam towards one-bond coordinates, with the identity of the bond depending on the sign of the logarithm of $\zeta$. As shown in the orthogonal top view, the partition of the intersection seams by the sign of $v$ is maintained. *The effect of varying $\zeta$ is to "sweep out" the area available to the seam as granted by the necessary conditions given in eqns. 37.*

## IX. $S_2/S_1$ Intersections occur near the Franck-Condon point when the coupling between the ends is the same as their coupling to the bridge.

Let us carefully inspect Figs. 11 and 9: when $\xi=1$, the $S_2/S_1$ intersections from diagonally adjacent cells in the Torsion Plane "collide" and switch direction. When $\xi<1$, these branches of the intersection seam proceed outward from the centre of the cell towards opposite edges along a disrotatory line in the plane. For $\xi>1$, they proceed inward along a conrotatory coordinate. The $S_2/S_1$ branches collide at $(\theta_L,\theta_R)=(0,0)$ – that is, at the Franck-Condon Geometry.

An analogous phenomenon also occurs for more general choices of the parameters, when neither $\xi$ nor $\psi$ necessarily equal zero. Figure 12 displays intersection seams at three values of $\xi+\psi$, and there are $S_2/S_1$ intersections at planar configurations of the model when $\xi+\psi=1$. These intersections occur *at all values of* *the affine parameter* $\frac{\xi-\psi}{\xi+\psi}$. The occurrence of these intersections is associated with a change in the structure of the $S_2/S_1$ seam. In the figure, the $S_2/S_1$ intersection curves "downward" when $\xi+\psi < 1$, but "upward" when $\xi+\psi > 1$.

The change in the behavior of the $S_2/S_1$ seam has important consequences for where the intersections lie in the Torsion Plane. When $\xi+\psi < 1$, the $S_1/S_0$ intersections and the $S_2/S_1$ intersections lie along perpendicular lines in the plane, but when $\xi+\psi > 1$, they lie along the same line. This clearly changes the geometry of pathways for deactivation following excitation, because when both $S_1/S_0$ and $S_2/S_1$ intersections lie on the same line, a straight-line trajectory in the plane can lead to complete $S_2$ deactivation. Also, identical straight-line trajectory following excitation to $S_1$ could "upfunnel" and become nonadiabatically trapped on the $S_2$ surface. Such behavior has been suggested as the origin of excitation-dependent broadening in the excitation spectrum of the chromophore of photoactive yellow protein[65]. On the other hand, if the intersections lie on perpendicular lines, then no straight line path from the planar configurations can lead to complete $S_2$ deactivation.

When $S_2/S_1$ intersections occur at the Franck-Condon Geometry as a consequence of the collision of the seam with itself, these intersections are not conical, but glancing ("Renner-Teller" intersections). Deviations in the value of $\zeta$ cause the intersection branches to miss each other, restoring the conical character of the intersection. These interesting characteristics of the $S_2/S_1$ intersection seams are highlighted in Figure 13.

**X. If the coupling is the same along con- and dis-rotatory coordinates, several $S_2/S_1$ intersections occur.**

As the total left-right coupling strength rises above the bridge-end coupling, the $S_2/S_1$ intersection "turns over", as highlighted in Fig. 12. An interesting consequence of this is that when the affine parameter is close to zero, the $S_2/S_1$ intersection branches will cross the Torsion Plane multiple times. Figure 14 displays this behavior, and shows that when these conditions hold, the $S_2$ surface is "tight against" the $S_1$ surface. For larger values of the affine parameter, the $S_2$ and $S_1$ surfaces avoid each other more effectively - particularly in regions of high bridge twist.

The occurrence of multiple $S_2/S_1$ intersections in a unit cell of the Torsion Plane should have profound consequences for the dynamics, because there will be more ways for the motion of the molecule to induce internal conversion from $S_2$ to $S_1$ (or "upfunnelling" into $S_2$). Furthermore, because the energetic avoidance between $S_2$ and $S_1$ is small over large regions of the entire plane, nuclear dynamics occurring on these surfaces may be highly coherent, depending on the precise value of the characteristic energy scales for electronic and nuclear motion.

**XI. When the coupling to the bridge is different for different ends, the $S_1$ surface favors twisting one bond more than the other.**

Our model predicts that when the coupling of the bridge to the different ends is different, the $S_2$ and $S_1$ surfaces bias will bias the motion towards different directions in the Torsion Plane. We highlight this behavior in Figure 15. This behavior is easy to understand – the intersections between the $S_2$ and $S_1$ surfaces will usually occur at "high elevation" on the $S_1$ surface, and at "low elevation" on the $S_1$ surface. This induces curvature on the $S_1$ surface, which directs motion away from the $S_2/S_1$ intersection, while trajectories that pass through the intersection from $S_2$ may end up on the other side. When the $S_2/S_1$ intersections occur off of the con- or disrotatory lines in the plane (i.e. when $\zeta \neq 1$), then the curvature of the $S_1$ surface at the planar configurations will bias dynamics so that torsion about one of the bonds is greater than the other. The curvature of the $S_2$ state will be biased in a complementary way. In this fashion, we expect that *excitation into different states may lead to photoisomerization about different bonds*.

### XII. Intersections separate regions of distinct charge localization.

An interesting physical point made by our model is that the intersection seams tend to separate regions of distinctly different distributions of fragment population. There are a few distinct regimes, identifiable by qualitatively different population distributions over the torsion plane. For symmetric bridge-end coupling, qualitative changes in the population distribution over the torsion plane are seen in the vicinity of $\xi+\psi \sim 1$. The overall behavior is captured in Figure 16, which shows charge distributions at several values of $\xi$ for constant $\psi=0.5$, and shows that the charge distribution changes abruptly at the intersections. When $\xi+\psi<1$, the $S_0$ population is evenly distributed at the planar configurations (FC), the $S_1$ populations are distributed between the two ends, and the $S_2$ population is concentrated on the bridge. As shown, the charge distribution is twist-dependent in all three adiabatic states, and changes most abruptly near the intersections impinging on any given state.

The population distribution over the fragments can be viewed as a measure of charge localization/delocalization. The dramatic change in charge localization near the intersections could have significant consequences for the dynamics of photoisomerization in condensed phase environments. Model studies on simple double-bonded systems suggest that the response of the solvent in such cases can lead to different dynamical pathways according to the effective mass of the solvent coordinate[58], and simulations of photoisomerization in methine dyes in water clusters suggest that the charge-transfer nature of the intersection seams induces environmental dependence in the decay times.[45]

**XIII. Biased bond twisting coincides with biased charge localization.**

Connecting the points of the last two sections leads to an interesting picture of the dynamics that can occur on the $S_1$ state. The distortion of the $S_1$ state that occurs when the bridge-end coupling is asymmetric leads to biasing one bond over the other, by the curvature of the surface, near the Franck-Condon region. This behavior is naturally related to the location of the intersections impinging on the $S_1$ state, because $S_2/S_1$ intersections lie at high elevation on the $S_1$ surface, while $S_1/S_0$ intersections lie at low elevation on the same surface. In the previous section, however, we have shown that the location of the intersections is also relevant in the determination of the charge distribution.

Figure 17 shows that this relationship of the population distribution and the potential surface curvature to the intersections can be transitive – the distortion of the $S_1$ potential energy surface is linked to the charge localization character of the $S_1$ state. When the bridge-end coupling is symmetric, the curvature of $S_1$ near FC is equivocal with respect to the two single-bond twists, and also equivocal with respect to charge localization on the left and right sides. Twisting off of the con- and disrotatory lines leads to charge accumulation on one end. *If the coupling is asymmetric, then the molecule is polar even in the Franck-Condon region.* If the molecule twists in one direction, this polarity will be maintained, but if it twists in the other direction, the polarity will eventually switch. The "critical twist", where the switching occurs, will depend upon the magnitude of the asymmetry.

For a dye in solution, the interaction of the dye with the solvent will often have a significant electrostatic component. This is intuitively true in polar solvents, but may also be observed in solvents that are not normally considered polar[66]. Under this type of interaction, the charge localization properties of the photoisomerization pathways may be very important to the dynamics. It is possible that the interactions could cause symmetry breaking in a chemically symmetric dye, for example, and one might expect to see a separation of timescales for different pathways, similar to what one might expect of an asymmetric dye. Figure 17 seems to suggest that the effects of biasing by potential curvature and biasing by charge-localization may be mutually reinforcing, leading to enhanced selectivity for the condensed-phase situation relative to the molecule in isolation.

**XIV. Discussion**

In this paper, we have examined the intersection seams within a very simplified orbital overlap model of a degenerate monomethine dye. We have pointed out several observations, which are relevant to recent work on such systems. Here, we expound on these connections.

Within our model, the $S_0/S_1$ intersection seam is *always* confined to a limited region of high twist, by conditions of necessity expressed in (37). This conflicts, *prima facie* with the suggestion, put forward to explain the fast decay component of 1122C[35], that there is an accessible seam of conical intersection close to the Franck-Condon point. No such intersection can occur in our model for any value of the parameters. Preliminary work by us on generalized versions of the model, with the degeneracy lifted, shows that this is also true in the general case. The existence of an intersection seam close to the FC point was suggested based upon nonadiabatic simulations of small model *tri*methine *strepto*cyanine[12]. Comparison of such a molecule with the molecules in Figure 1 is interesting, because it will be both more and less complicated. A trimethine bridge will have more degrees of freedom. It is possible that the additional degrees of freedom may lead to conical intersections near the Franck-Condon point. – perhaps by allowing a given twist to be distributed over more bonds. Streptocyanines are cyanines where the end groups are not rings, but simple amines. If nonplanar distortions of the amine group allow strong σ-π mixing (through facile pyramidalization, for example), then our simple model may not apply.

Although the $S_1/S_0$ intersections in our model are confined to regions of high bridge twist, *the $S_2/S_1$ intersections are not*. These intersections can arise at planar or nearly planar geometries, and can occur *at* the Franck-Condon point itself. There is a two-photon excitable higher excited state in several monomethine dyes, excitation of which gives rise to emission from the same state as as for the one-photon case[67,68]. For dyes under high pressure, the fluorescence intensity can differ, suggesting that the path to the emitting state following one and two-photon excitation may be different[67]. In the monomethine dye Malachite Green, the $S_2$ state can be accessed by both one- and two-photon excitation, and emission can be detected from both states[69,70]. Decay to the $S_1$ state occurs on ultrafast timescales and the dynamics suggest torsional motion may be involved. *Ab initio* models of photoactive yellow protein chromophores suggest that $S_2/S_1$ intersections may directly[65] or indirectly[71] affect the photophysics of these systems. Our model is broadly consistent with all of these phenomena. It suggests that internal conversion to the $S_1$ state should be possible, that it should occur at in regions with low twist (consistent with radiative outcomes), and that limited twisting motions may be involved.

Our model highlights the importance of con- and or dis-rotatory motion to the internal conversion of monomethine dyes. In the limit of a perfectly symmetric dye, intersections only occur along these coordinates. Disrotatory motion is sometimes referred to as *hula-twist* in the photoisomerization literature[72,73], where it is put forward as a volume-conserving pathway, invoked to explain photoisomerization in restricted environments. It has been explicitly invoked several times in the context fluorescent protein chromophores[28,62,74]. Our results suggest that steric hindrance is not necessary as a precondition to decay by this pathway, which may be quite natural for dyes with balanced end groups.

Our model suggests that when the coupling of end groups to the bridge is *not* symmetric, dynamics occurring on the $S_1$ state will be biased towards one bond over the other, and the biasing on $S_1$ and $S_2$ will be complementary. This is an interesting result in itself, because it suggests how the different bond torsions can be independently addressed in laser control experiments. It also suggests that the dynamics under linear excitation of an asymmetric dye may not be as complicated as we thought. As the dye becomes progressively more asymmetric, one pathway will dominate the $S_1$ dynamics. We think that this physics may be responsible for photoswitching in reversibly photoswitchable fluorescent proteins, where the switching mechanism involves a concerted change of isomeric and protonation state of the chromophore. There have been several recent publications which cite bond-selective photoisomerization in the photoactive yellow protein (PYP) chromophore[71,75,76] and fluorescent protein (FP) chromophores[42-44,62,77]. PYP and FP chromophores are oxonol methine dyes. We suspect that the reported behavior of both PYP and GFP chromophores have their origin in the same physics, and our model captures important parts of this physics. Similar behavior arises in Hückel models of polymethine cations[78], although the deep connection between these studies and the photophysics of the PYP and FP chromophores may not have been realized.

The same physics that leads to different bond selectivity on the $S_1$ and $S_2$ states of our model may also lead to a separation or broadening of timescales for decay from the $S_1$ state itself. The motion of the $S_2/S_1$ intersection, which is at high elevation on $S_1$ and low elevation on $S_2$, influences the curvature on the $S_1$ surface. When the intersections move off of the con/dis-rotatory lines towards regions of uneven twist, this is reflected at the topography of the Franck-Condon region. Our model is aggressively minimalistic; curvature distortions are always clearly connected to the intersections, which are always present. In a more highly dimensional situation, barriers may still be connect to the intersection seam, even if the intersection seam itself is inaccessible. Similar physics underlies "avoided crossing state", "perfectly resonating state" and "twin state" models of chemical reactivity[79-83].

Our model suggests that curvature-induced biasing of the $S_1$ surface may be reinforced by the charge localization, which also depends upon the location of the intersections impinging on $S_1$. Asymmetry in the coupling can lead to the induction of polarity even at the Franck-Condon point (see Fig. 17). This is important, because in a condensed phase environment, the charge localization on the molecule and the polarization of the environment will be linked through the reaction field into a feedback loop. This implies a certain degree of nonlinearity in the system, of the sort that could potentially lead to symmetry breaking *even if the dye is symmetric in isolation.*

This behavior hints at an alternative explanation for the separation of decay timescales seen in recent experiments on the symmetric dyes 1122C[35] and NK88[40]. Both of these dyes are symmetric, so nonlinear solvent interactions must be invoked to explain any asymmetry in the coupling. A situation is conceivable where the dye-solvent system may exist in a distribution of states so that the system is sometimes "pre-organized" for photoisomerization about one bond more than the other. A separation of timescales may be induced in the dynamics, with a fast timescale attributed to populations where the dye and solvent states are arranged "in sync", allowing decay at speeds rivalling the solvent reorientation timescale, and slower decay for populations where the solvent must reorient to solvate the appropriate charge-localized excited state.

Further work is needed to flesh out these ideas. Numerical models suggest that modulation of the coupling by the environment is likely to be dominated by tuning of the diagonal elements of the effective Hamilitonian[84,85]. Still, we would expect that follow-on effects would lead to tuning the off-diagonal elements, since the diagonal and off-diagonal terms are both linked through the molecular geometry, which, in turn, responds to the electronic state. This feedback loop is the physical origin of soliton behavior in long-chain polyacetylenes[86] and polymethine cations[87-89].

The intersection seams in our model are all *continuous and connected.* This is true for the entire seam, spanning both $S_1/S_0$ and $S_2/S_1$ branches. The reasons are straightforward. The left hand side of equation (29), which sets necessary and sufficient conditions for the intersection of states within the model, is a polynomial in the matrix elements, and the matrix elements are represented by functions which are continuous and single-valued modulo the periodicity of the angles $\theta_L$ and $\theta_R$. If $u$ and $v$ are single-valued and continuous, and the matrix elements on which they depend are also single-valued and continuous, then the intersection will be single-valued and continuous, consistent with a conjecture put forward by Coe, Levine and Martinez[74]. Similar considerations would also apply for intersection seams in a similar *ab initio* model based on a self-consistent field, as long as the matrix elements are continuous and single-valued. We postulate that this will be true so long as the underlying SCF solution varies continuously over the relevant region of configuration space. This is not guaranteed, however, because SCF solutions are generally not analytic with respect to dilation[90,91]. The continuity of the intersection seams will, in general, be limited by the analyticity of the SCF solution.

## XV. Conclusion

We have proposed a simple matrix model for the Hamiltonian of a degenerate monomethine dye, and described interesting physics, which arises when the intersection seams of the model are examined in detail.  We are continuing to develop more elaborate models along similar lines.  We will discuss these in later publications, in which we will refer to the present work as a useful reference case.  Even at the present level of simplicity, the model exhibits rich physics relevant to important topics of current interest in the photoisomerization of monomethine dyes

**Acknowlegement** This work was partially supported with funds from the Australian Research Council Discovery Project DP0877875.  We thank Anthony Jacko and Michael Smith for readings of the manuscript.  We additionally thank Anthony Jacko for bringing to our attention an error in the eigenvalues formulas. We acknowledge helpful discussions with Steven Boxer, Paul Brumer, Irene Burghardt, Dan Cox, Philippe Hiberty, Noel Hush, Todd Martínez, Stephen Meech, Ben Powell, Jeff Reimers, Fritz Schaefer, Shason Shaik, and Tom Stace.  Some of the graphics were generated with VMD[92].

**Figure Captions**

**Figure 1.** Examples of monomethine dyes. The molecules resonate between Lewis structures which invert bond alternation and redistribute the formal charge. Examples include, from top to bottom, the symmetric monomethine cyanine dye 1122C, the symmetric monomethine cyanine dye NK88, the asymmetric monomethine cyanine dye Thiazole Orange, and the chromophore of the green fluorescent protein, an asymmetric diarylmethine oxonol dye.

**Figure 2.** The 4 isomers of the asymmetrical dye Thiazole Orange, which differ by (Z,E) isomerism of the bridge. Thiazole Orange is the least symmetrical of the three example molecules in Fig. 1, and so all isomers are distinguishable. They are labelled according to usual organic chemistry nomenclature.

**Figure 3.** Schematic description of a conical intersection in the branching plane. Reactants (green lump) are promoted to the excited state surface by a photon, forming the Franck Condon State (red lump) which then evolves in in all possible ways (transparent orange lumps along black lines) on the excited state. Upon passing near a conical intersection between the surfaces, population can return to the ground state and continues to evolve to form one or more products (yellow lumps). The schematic highlights the possibility that the reaction does not complete – all end products are equally likely in this picture, so some evolutions will terminate in the reactant basin.

**Figure 4.** Three conceivable situations where the shape of surfaces, and their interaction, can influence photochemical pathways. On the left, two states are biased towards the same pathway and ensuing dynamics may converge to a common evolution. In the middle, the electronic states are close in energy at the Franck-Condon (FC) region, and are not biased strongly, so there is ambiguity in the pathways and no clear connection between the dynamics and the initially excited state. On the right, the states are biased differently, so that the dynamics on the different states diverge. Other situations are also conceivable.

**Figure 5.** Localized-orbital active space representations for monomethine dye systems. For every monomethine dye, there is a "methine adapted" three-orbital solution to the state-averaged complete active space self consistent field problem. Monomethine cyanine dyes (left) have a 2-electron solution, and diarylmethine dyes (right) possess a 4-electron solution. In either case, the many-electron state space is six-dimensional and has a natural valence-bond structure in the localized representation (bottom). The energetic ordering of the localized orbitals is inverted in the two dye classes.

**Figure 6.** An introduction to our model Hamiltonian. *(Top left)* Our model is a three by three diagonally degenerate matrix (which we can make traceless by a shift). *(Top right)* The interaction matrix elements are cosine functions of two angles, $\theta_L, \theta_R$, multiplied by functions of real positive numbers $\xi, \psi, \zeta$. *(Bottom right)* A toy model using simple p orbitals, where the overlaps between p orbitals at sites on the vertices of a triangle have the same functional dependence on $\theta_L$ and $\theta_R$ as the Hamiltonian elements.

**Figure 7.** Schematic representation of relevant coordinates in terms of the geometrical model in figure 4 (top) and as vectors in the $(\theta_L, \theta_R)$ plane (bottom). Single bond twists change one angle while leaving the other constant. The conrotatory and disrotatory twist coordinates are antisymmetric and symmetric combinations of the single bond twists,

respectively. The parity of the combination coordinates depends on the "handedness" convention used to define the torsion angles, which here uses left and right hand rules for left and right torsion angles. Changing the handedness of the definition for one of the bonds is equivalent to interchanging the parameters $\xi$ and $\psi$. In the context of an untwisted molecular frame with $C_{2v}$ symmetry, conrotatory twisting preserves $C_2$ symmetry and breaks $C_s$ symmetry, while the disrotatory twist breaks $C_s$ symmetry and preserves $C_2$ symmetry.

**Figure 8.** The analytic eigenvalues of a 3x3 symmetric matrix can be expressed by two parameters $u$ and $v$, which are polynomials of first and higher traces of the matrix. The space of all degeneracies between the eigenvalues can be expressed as a relationship between these two parameters. (Top) Eigenvalues plotted over a plane spanned by the parameters $u$ and $v$. The conditions $u < 0$ and $u^3+v^2 < 0$ (region shown at bottom) are sufficient to guarantee real eigenvalues of a 3 x 3 matrix and are equivalent to symmetry and positive definiteness of the matrix. When the inequality is strong, 3 non-degenerate eigenvalues exist. On the boundary (highlighted in black) at least 2 of the eigenvalues are degenerate. $S_1/S_0$ degeneracies occur on the $v < 0$ part of the boundary. $S_2/S_1$ degeneracies occur on the $v > 0$ region of the boundary, and a 3-state intersection occurs at $(u,v) = (0,0)$.

**Figure 9.** Intersection seams in the limit where $\psi=0, \zeta=1$. In this case, the model is three-dimensional. The intersections have be visualized by filling in a region defined by eqn. (31) with $\delta = 0.005$. The distinct branches are colored according to the sign of $v$ (see eqns. 33,34); red contours enclose $S_1/S_0$ intersections and yellow contours enclose $S_2/S_1$

intersections. The intersections in a unit cell of the ($\theta_L, \theta_R$) plane (left) are continued over the periodic boundaries as the cell is extended (top right). The $S_2/S_1$ intersection manifolds from diagonally adjacent unit cells collide and diverge at the corners, changing direction when $\xi = 1.0$. At $\xi = 0.0$, the model becomes singular at ($\pi/2, \pi/2$), giving rise to a 3 state intersection where both $v$ and $u$ (eqns 35,36) vanish. As $\xi$ becomes infinite, the $S_2/S_1$ and $S_1/S_0$ intersections will asymptotically approach each other at the points ($\pi/4, \pi/4$) and ($3\pi/4, 3\pi/4$). The limit where $\xi=0, \zeta=1$ would look identical in a plot of the ($\psi, \theta_L, \theta_R$) space, but rotated by $\pi/2$.

**Figure 10.** The $S_1/S_0$ intersections seams in the model are always confined to a limited region of high bridge twist. (Left) Intersection seams for a collection of values of $\xi+\psi$ and $\zeta$. The intersections are visualized by eqn. (31) with $\delta=0.004$. The $S_1/S_0$ intersections always occur in the shaded region, and $S_2/S_1$ intersections occur outside. The shaded region corresponds to the sign of $v$ (eqns 33,34). The region where $v<0$ is limited to regions of high bridge twist for all values of $\xi, \psi$ and $\zeta$, but only depends explicitly on the ratio of $\xi-\psi/\xi+\psi$. In the context of our toy molecular model (Fig. 6), this is the difference between the $\pi$ and $\sigma$-type components of the left-right coupling relative to their sum.

**Figure 11.** (Top) Changes in intersection seams within the model corresponding to deviations of $\zeta$ from 1 in the limit where $\psi = 0$. In this limit, intersections can be found along conrotatory lines ($S_1/S_0$ and $S_2/S_1$ intersections) and disrotatory lines ($S_2/S_1$) intersections in the plane. The case for $\zeta = 1$ (centre) is identical to the case

shown in Fig 9. Deviations from $\zeta = 1$ result in movement of the intersection seams off of the lines corresponding to con- or disrotatory torsion, biasing the twist distribution towards one-bond torsion coordinates. The intersections will approach distributions where one angle is equal to $\pi/2$ in the limts where $\zeta \to 0, \infty$. For intermediate values of $\zeta$, intersections will occur along lines with mixed but biased twist distribution. (Bottom) Top views (orthogonal projection) of the plots superimposed over a partitioning of the angle-angle plane into regions where $v > 0$ (darker) and $v < 0$ (white). The effect of changing $\zeta$ is not to change partitioning, but to allow the seam to sample the region allowed in principle by the sign of $v$ (eqns. 33,34).

**Figure 12.** $S_2/S_1$ intersections can occur at planar geometries (i.e. near the Franck-Condon Geometry) when the coupling between the ends, and the coupling of either to the bridge, have the same magnitude (i.e. when $\xi+\psi \sim 1$). (Top) $S_1/S_0$ (red) and $S_2/S_1$ (yellow) intersection seams at constant $\xi+\psi=0.75$(left) ,*1.0*(center) ,*1.25*(right). When $\xi+\psi= 1$, there are $S_2/S_1$ intersections at the corners of the unit cell for *all values of* $\frac{\xi-\psi}{\xi+\psi}$. This is accompanied by a change in the curvature of the $S_2/S_1$ seam. (Bottom) The location of the $S_2/S_1$ seam in the Torsion Plane spanned by the angles $(\theta_L, \theta_R)$ at different values of $\xi+\psi$. When $\xi+\psi< 1$, the $S_2/S_1$ and $S_1/S_0$ intersections lie along perpendicular lines in the plane. When $\xi+\psi> 1$ this behavior changes, and both $S_1/S_0$ and $S_2/S_1$ intersections lie along a single line. Black arrows indicate the direction of motion of the seams with increasing $\xi+\psi$. Intersections were visualized using eqn. (31), with different branches colored according to the sign of $v$ (eqns. 33,34).

**Figure 13.** This figure focuses on a peculiar aspect of the $S_2/S_1$ intersections which occur at planar geometries (as in Fig. 12 center). (Top,Centre) When the coupling of the bridge to the ends is the same for both ends ($\zeta=1$), the seam occurs along a line in the plane described by $\theta_L-\theta_R=0$. (Top, left and right) When it is asymmetric $\zeta\neq1$, the seams originating from diagonally neighboring "unit cells" miss each other, and there is no discontinuity in the tangents to the seam. (Bottom) The collision of the degeneracy seams when coupling is symmetric ($\zeta=1$) results in a "glancing" (Renner-Teller) intersection of the potential surfaces (bottom center), in contrast to conical (Jahn-Teller) intersections, which arise when the coupling of the bridge to the ends is asymmetric ($\zeta\neq1$).

**Figure 14.** If the coupling between the ends is larger than their coupling to the bridge ($\xi+\psi\geq 1$) but the difference between con- and disrotatory coupling parameters is small ($|\xi-\psi|\sim0.1$), there are $S_2/S_1$ intersections, which occur at different twists. In this range of the parameters, $S_2$ is drawn tight over $S_1$, and the states are close over the large regions of the torsion plane (top right). However, the surfaces avoid each other more effectively as either con - or dirotatory coupling dominates the other ($|\xi-\psi|\sim\xi+\psi$). The number of $S_2/S_1$ intersections drops to two, modulo the periodic boundary conditions.

**Figure 15.** This figure illustrates how excitation-dependent selection of different bond torsions can occur in the $S_1$ and $S_2$ states of the model. (Right) Degeneracy seams (visualized in ($\xi$-$\psi$,$\theta_L$,$\theta_R$) space show that the seams occur along con- and disrotatory lines in the plane when the couplings of both ends to the bridge are the same ($\zeta$=1, top left), and move off of these lines when the coupling is asymmetric ($\zeta \neq 1$, bottom left). (Left) When $\zeta \rightarrow 1$, the curvature at the Franck-Condon point is equivocal with respect to torsion of the different bonds (top right). When $\zeta \neq 1$, the curvature is tighter for one bond than the other, which will bias the dynamics on the surfaces. The biasing of $S_2$ and $S_1$ is complementary.

**Figure 16.** Relationship between conical intersection seams (top), and charge-localization in the model. *(Left)* Conical intersection seams are shown for varying $\xi$ at constant $\psi$=0.5, $\zeta$=1. Cross sections are shown at $\xi$=0.25,0.625,0.75 and 1.0. *(Right)* Population (absolutely squared amplitude) of the fragment diabatic sites are plotted over the Torsion Plane. Populations for the states have been used as convex coordinates of an RGB color map. Areas where the cross-sections intersect the neighboorhood of the seam (eqn. 31, $\delta$ = 0.001) are shown by filled red ($S_1/S_0$) and yellow ($S_2/S_1$) regions. Charge localization is generally twist-dependent, and regions of different localization on the $S_1$ state are separated by intersections impinging on that state.

**Figure 17.** Synergistic relationship between energetic biasing of different bond torsions (top) and twist dependence of the charge localization (bottom). *(Top)*

Biasing in the $S_1$ potential energy surface is introduced by asymmetric bridge-end couplings. When the coupling of the bridge to both ends is symmetric *(top center)*, the curvature at the Franck-Condon point is equivocal with respect to both single-bond twists. Introducing asymmetric coupling alters the curvature to favor one bond over the other *(top left & right)*. *(Bottom)* Introducing asymmetric bridge-end coupling also exerts changes on the charge distribution over the fragments and it's dependence on the twist. When the coupling is symmetric, charge localization follows the twist distribution, for con-and dis-rotatory twists the charge is spread over the left and right fragments, but for single-bond twists it localizes on one side or the other, with twist-dependent polarity *(bottom center)*. When the coupling is not symmetric, the charge distribution at FC is polar. Twisting one bond does not change the polarity. Twisting the other does, but significant twist may be required to do so, depending on the magnitude of the asymmetry. Note that the cells of the Torsion Plane shown here are offset by $\pi/2$ relative to those in Fig. 16.

**Figure 1**

1,1'-DiEt-2,2'-Cyanine (1122C)

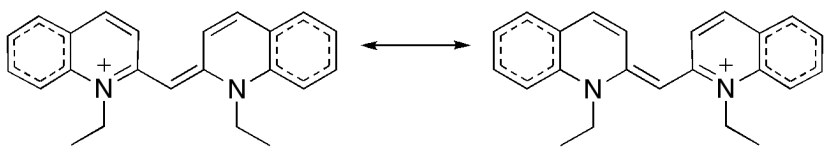

3,3'-DiEt-2,2'-Thiacyanine (NK88)

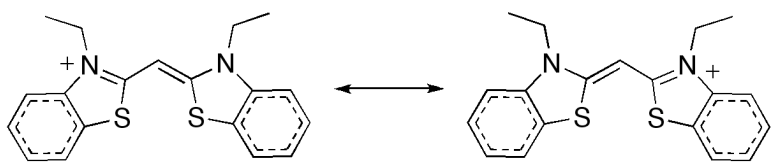

Thiazole Orange

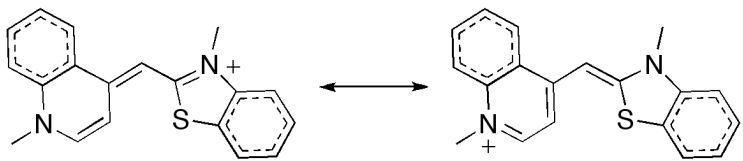

Green Fluorescent Protein Chromophore

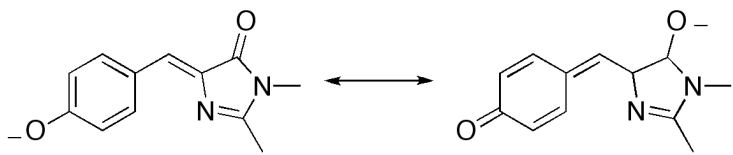

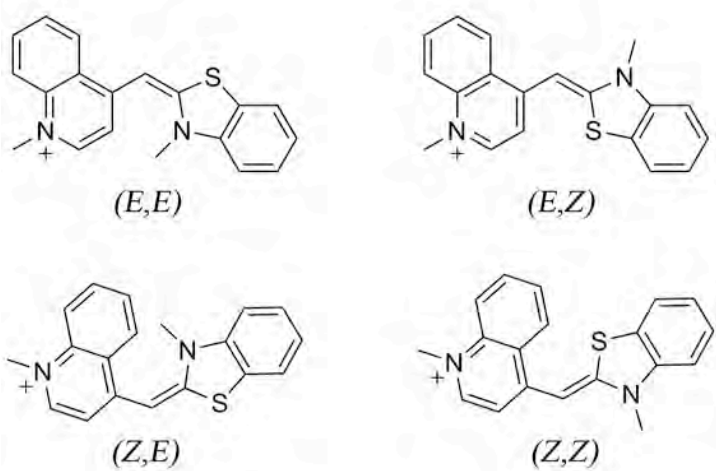

**Figure 2**

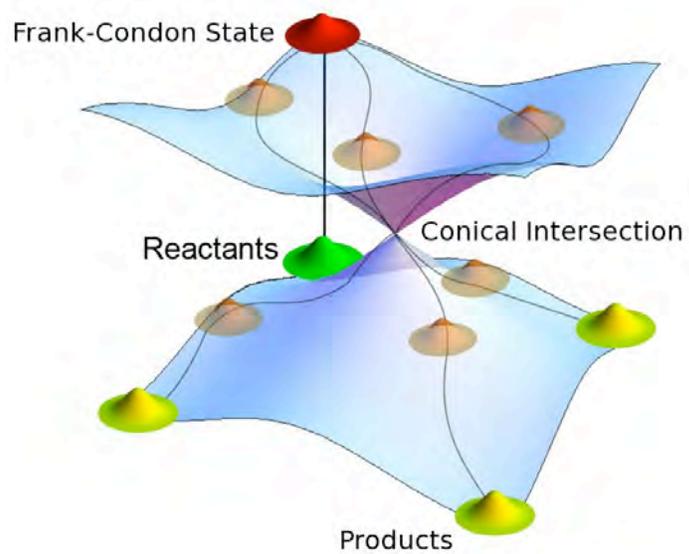

**Figure 4**

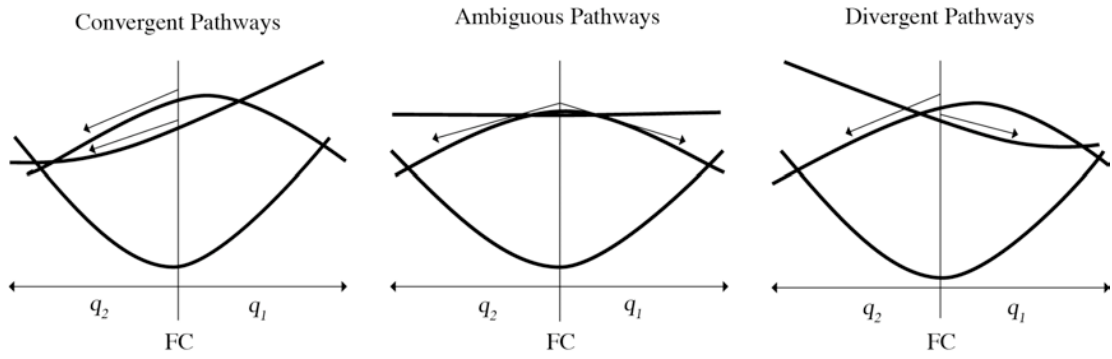

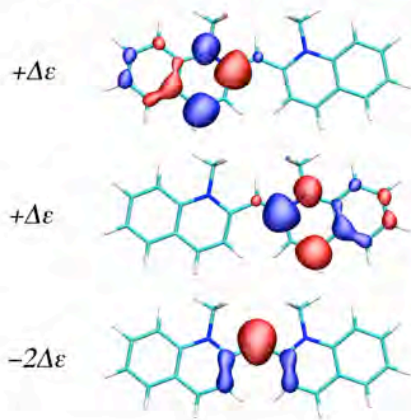
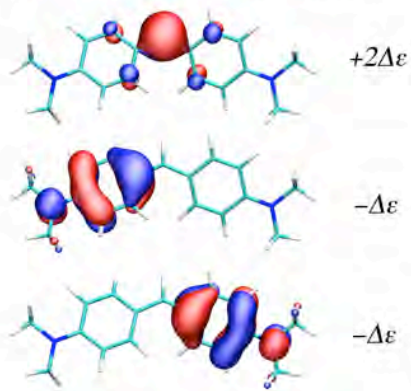
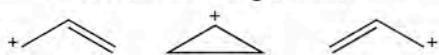
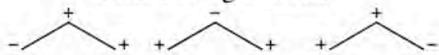
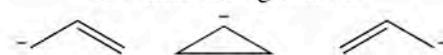
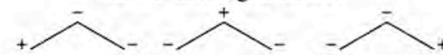

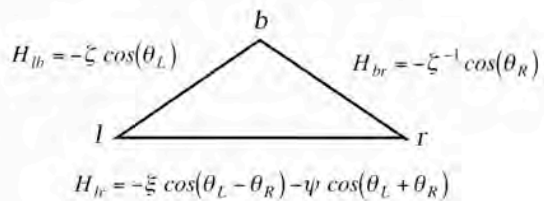
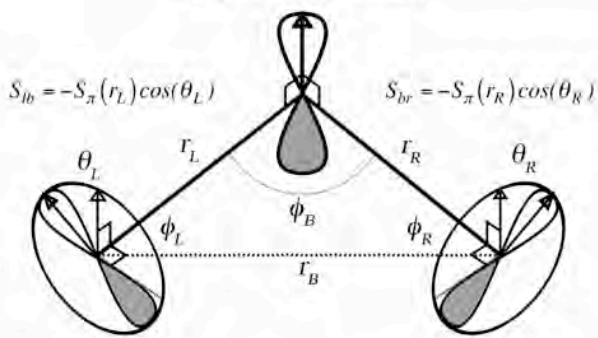

Figure 6

**Figure 7**

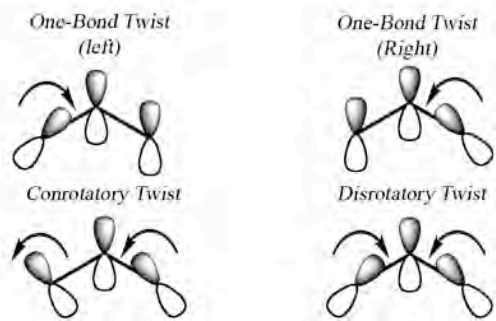

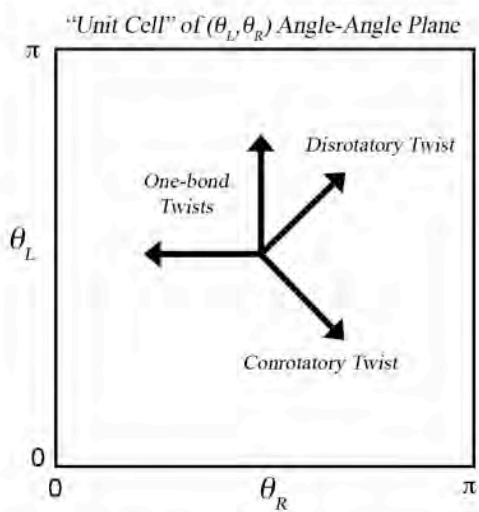

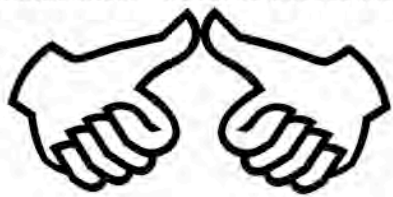

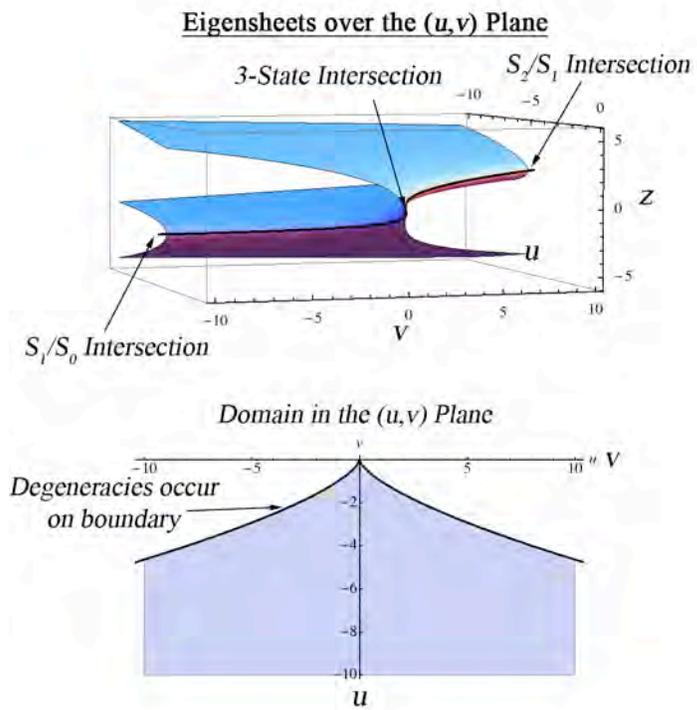

**Figure 9**

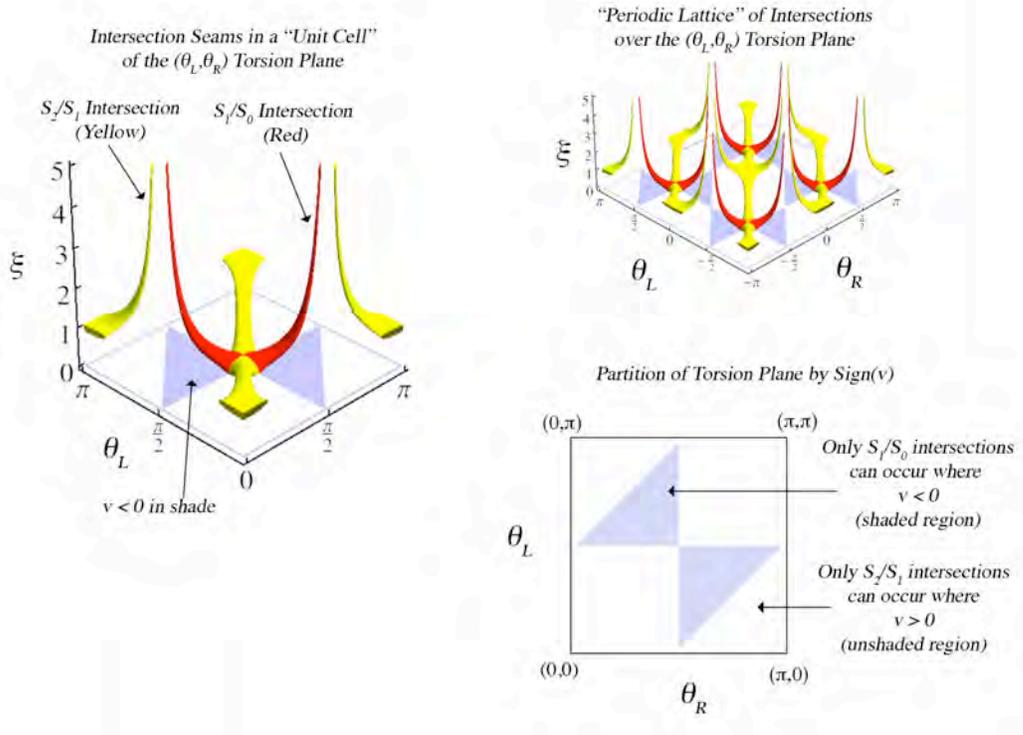

**Figure 10**

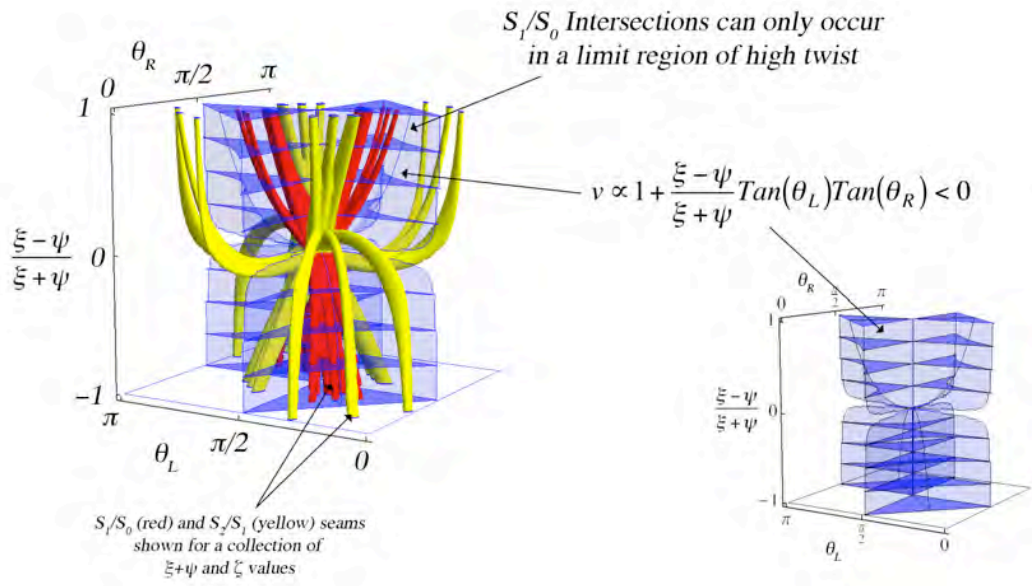

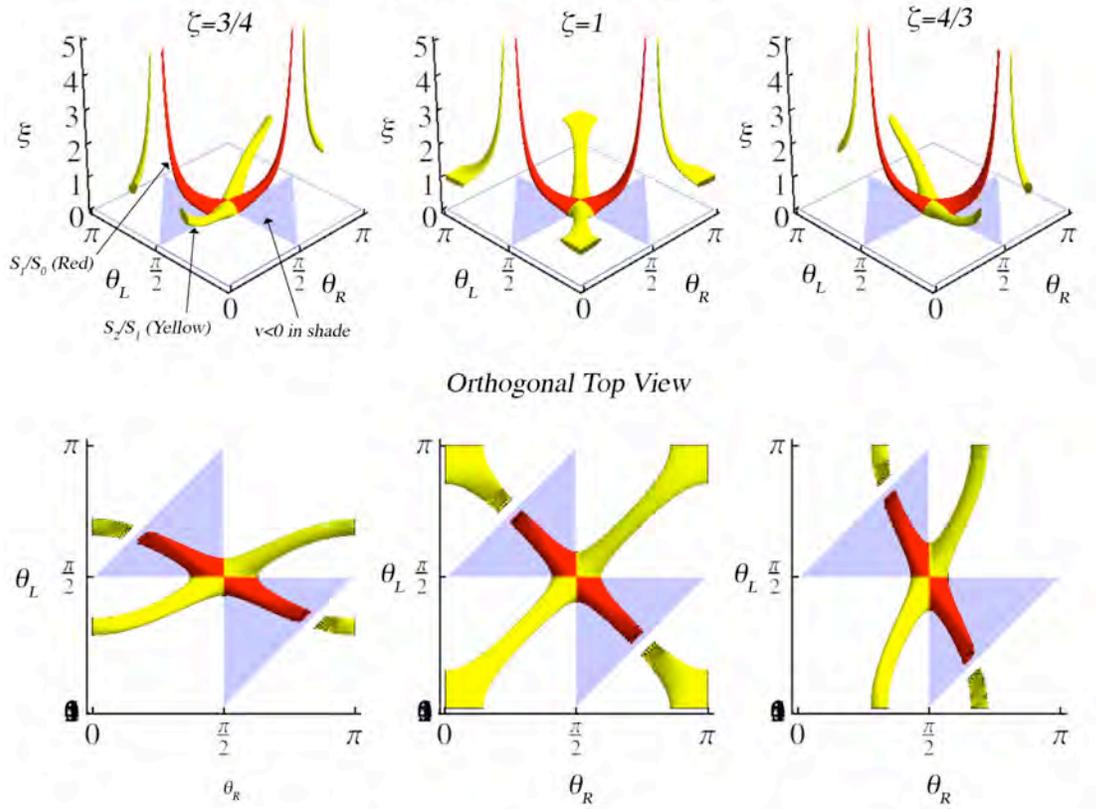

**Figure 12**

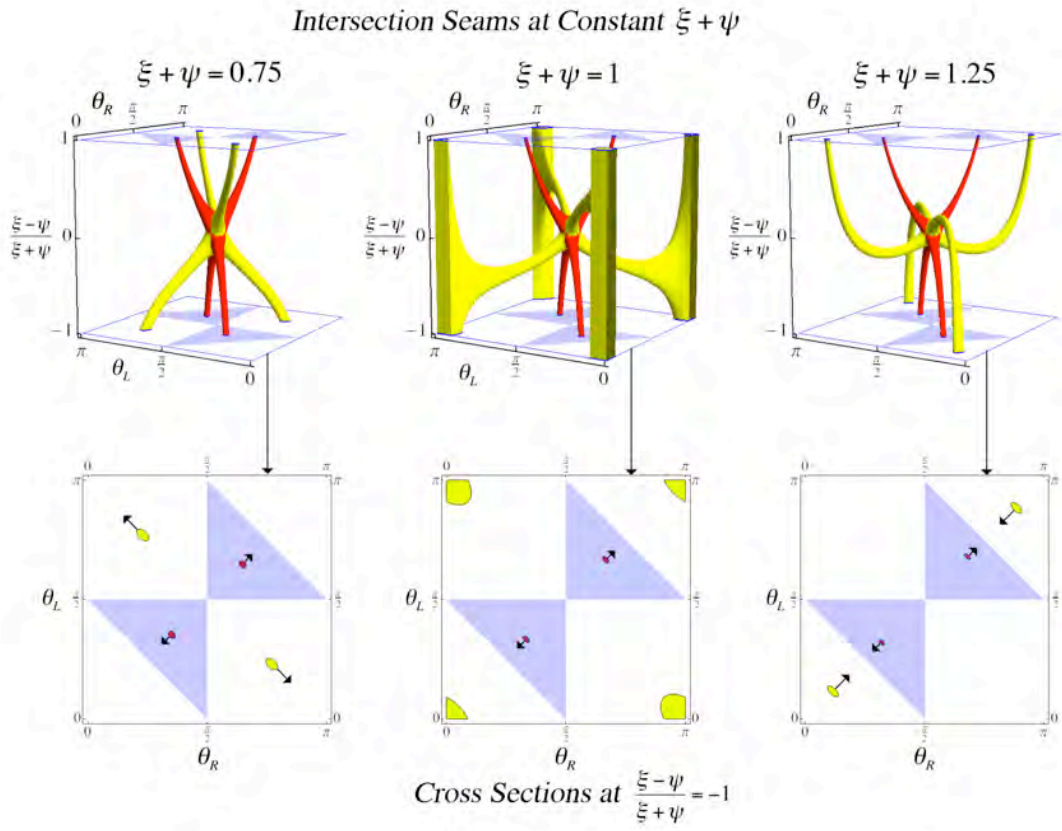

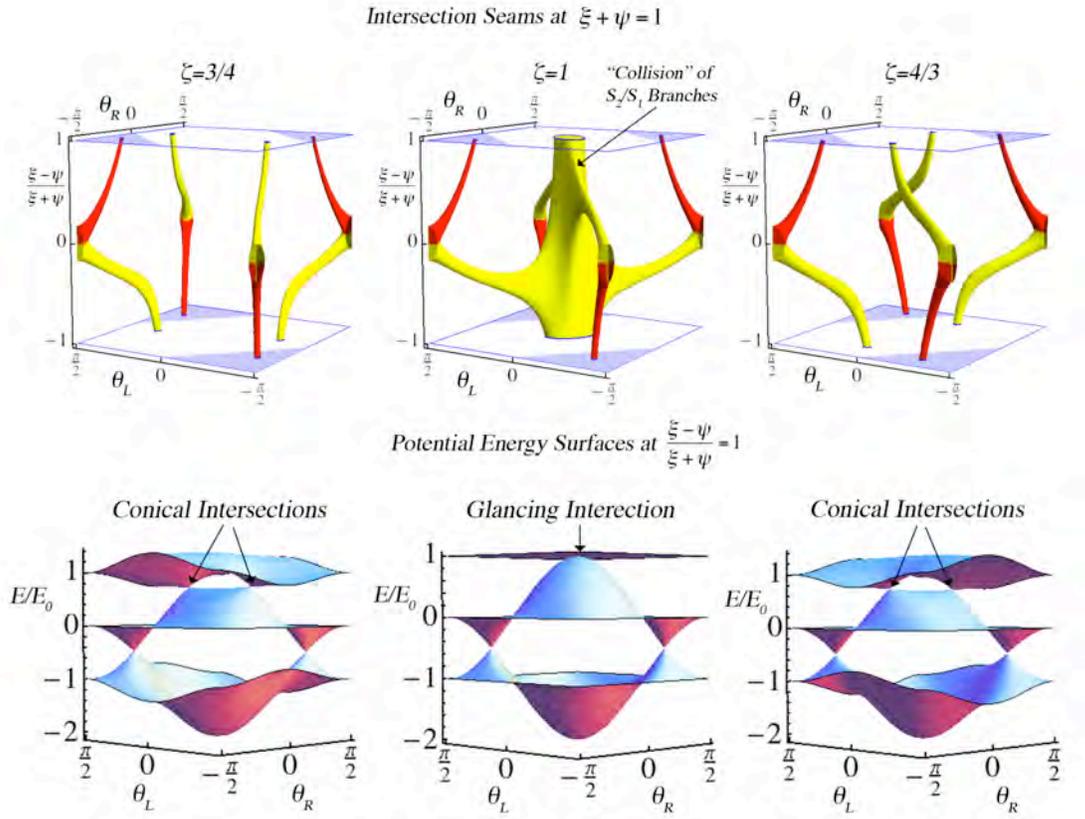

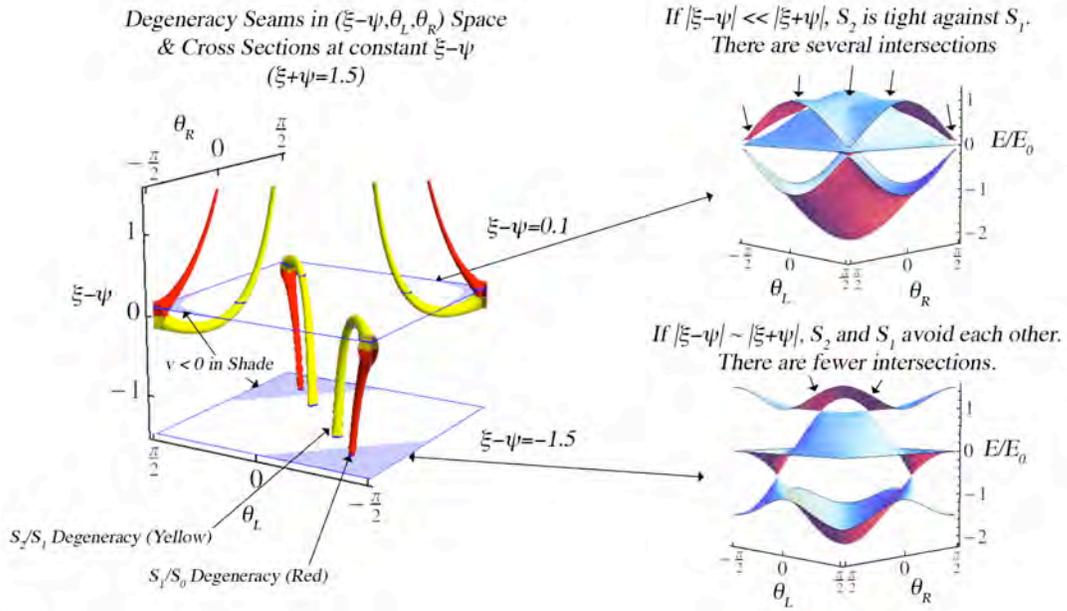

**Figure 15**

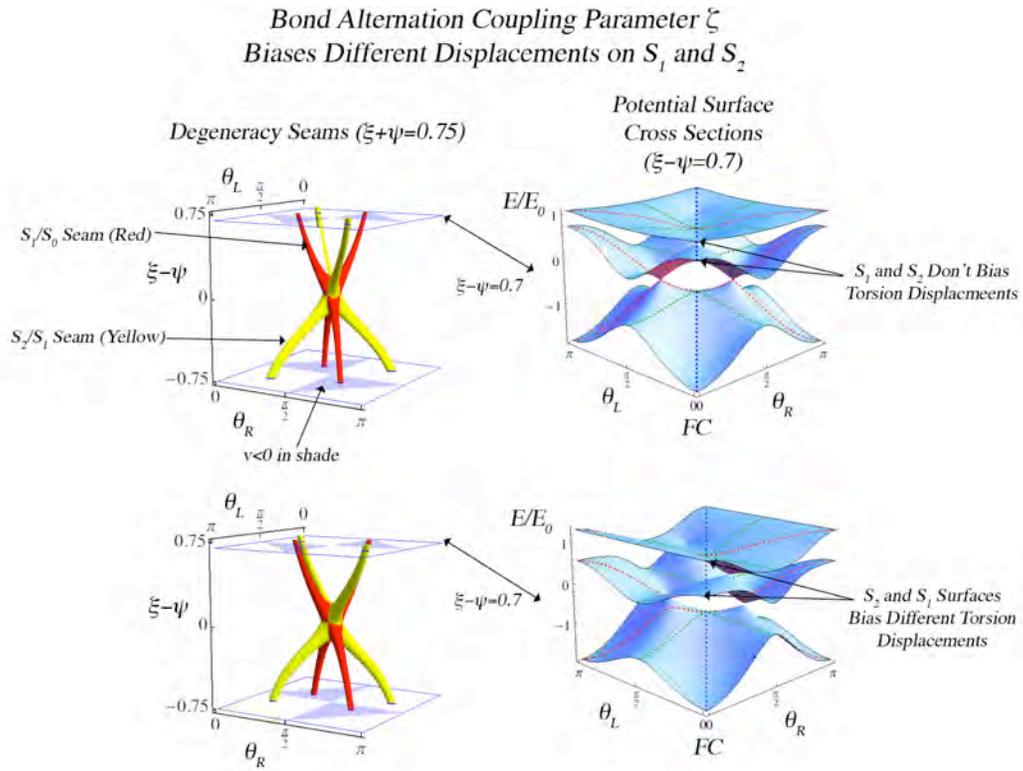

**Figure 16**

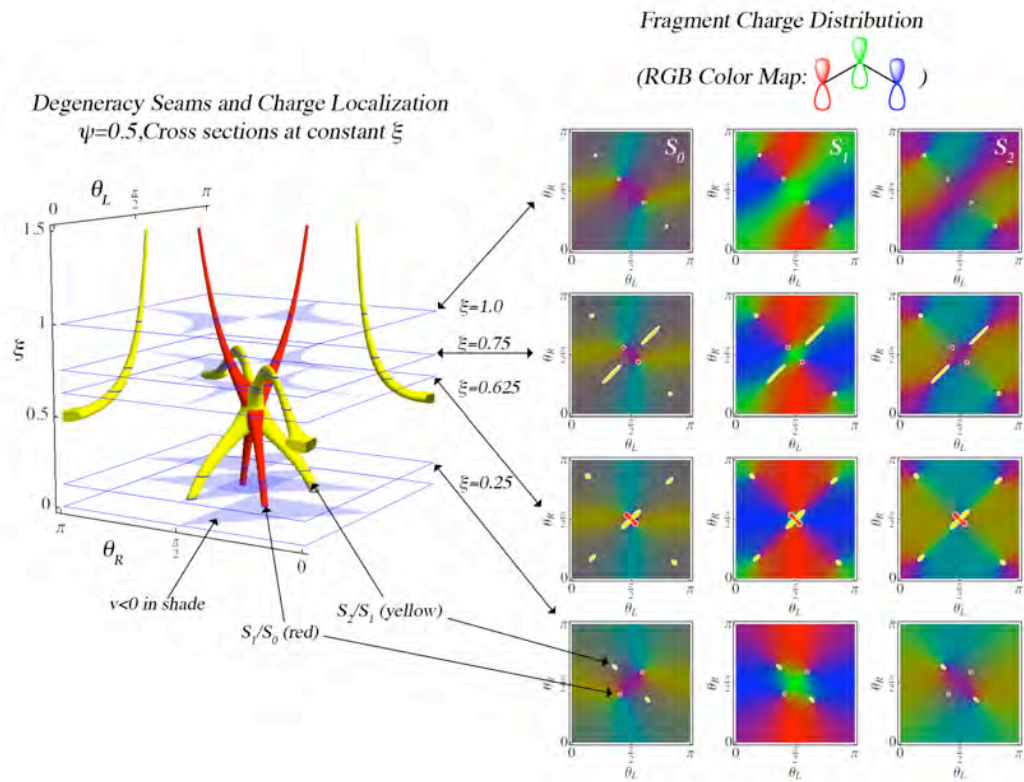

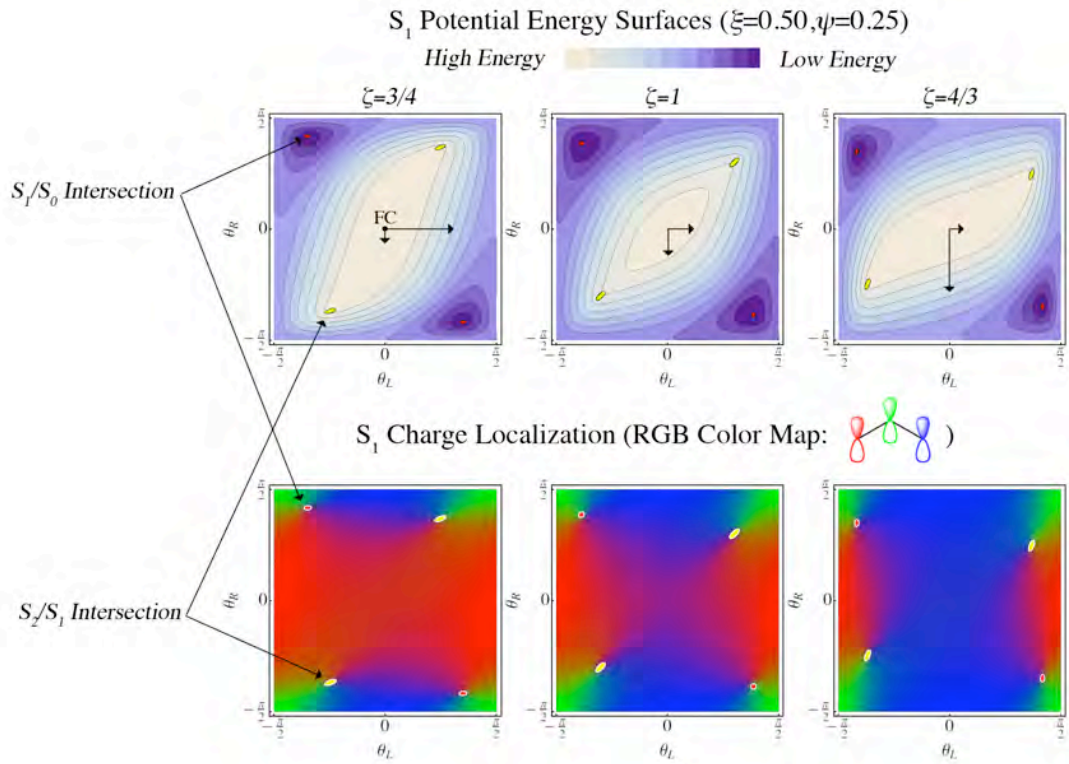